%% file: qbd.tex
\PassOptionsToPackage{table}{xcolor}
\documentclass[format=acmsmall, review=false, screen=true, anonymous=false]{acmart}

\usepackage{booktabs} 
\usepackage[ruled]{algorithm2e} 
\usepackage{natbib}
\usepackage{enumitem}
\usepackage{tabularx}
\usepackage{siunitx}
\usepackage{multirow}
\usepackage{subcaption}
\usepackage[justification=centering]{caption}
\usepackage{titlesec}
\usepackage{xcolor}

\definecolor{lavenderblue}{rgb}{0.9, 0.9, 0.98}
\newcolumntype{Y}{>{\centering\arraybackslash}X}

\acmJournal{PACMHCI}
\acmVolume{5}
\acmNumber{CSCW1}
\acmArticle{94}
\acmYear{2021}
\acmMonth{4}
\acmPrice{15.00}
\acmDOI{10.1145/3449168}

\setcopyright{acmlicensed}

\received{June 2020} 
\received[revised]{October 2020}
\received[accepted]{December 2020}

\begin{document}
\title[Supporting Serendipity]{Supporting Serendipity: Opportunities and Challenges for Human-AI Collaboration in Qualitative Analysis}

\author{Jialun Aaron Jiang}
\affiliation{
  \institution{University of Colorado Boulder}
  \department{Department of Information Science}
  \streetaddress{1045 18th St.}
  \city{Boulder}
  \state{CO}
  \postcode{80309}
  \country{USA}}
\email{aaron.jiang@colorado.edu}

\author{Kandrea Wade}
\affiliation{
  \institution{University of Colorado Boulder}
  \department{Department of Information Science}
  \streetaddress{1045 18th St.}
  \city{Boulder}
  \state{CO}
  \postcode{80309}
  \country{USA}}
\email{kandrea.wade@colorado.edu}

\author{Casey Fiesler}
\affiliation{
  \institution{University of Colorado Boulder}
  \department{Department of Information Science}
  \streetaddress{1045 18th St.}
  \city{Boulder}
  \state{CO}
  \postcode{80309}
  \country{USA}}
\email{casey.fiesler@colorado.edu}

\author{Jed R. Brubaker}
\affiliation{
  \institution{University of Colorado Boulder}
  \department{Department of Information Science}
  \streetaddress{1045 18th St.}
  \city{Boulder}
  \state{CO}
  \postcode{80309}
  \country{USA}}
\email{jed.brubaker@colorado.edu}

\renewcommand{\shortauthors}{Jialun Aaron Jiang et al.}

\begin{abstract}
Qualitative inductive methods are widely used in CSCW and HCI research for their ability to generatively discover deep and contextualized insights, but these inherently manual and human-resource-intensive processes are often infeasible for analyzing large corpora. Researchers have been increasingly interested in ways to apply qualitative methods to ``big'' data problems, hoping to achieve more generalizable results from larger amounts of data while preserving the depth and richness of qualitative methods. In this paper, we describe a study of qualitative researchers' work practices and their challenges, with an eye towards whether this is an appropriate domain for human-AI collaboration and what successful collaborations might entail. Our findings characterize participants' diverse methodological practices and nuanced collaboration dynamics, and identify areas where they might benefit from AI-based tools. While participants highlight the messiness and uncertainty of qualitative inductive analysis, they still want full agency over the process and believe that AI should not interfere. Our study provides a deep investigation of task delegability in human-AI collaboration in the context of qualitative analysis, and offers directions for the design of AI assistance that honor serendipity, human agency, and ambiguity.
\end{abstract}

%
%

\ccsdesc[500]{Human-centered computing~Collaborative and social computing theory, concepts and paradigms}
\ccsdesc[500]{Human-centered computing~Collaborative and social computing design and evaluation methods}

%
%


\keywords{Qualitative research; interview; AI; human-AI collaboration}

\maketitle

\input{qbd-body}

\end{document}

%% file: qbd-body.tex
\section{Introduction}

\begin{quote}
``You've got 500 codes first and then you've got notes everywhere all over them.'' (P05)
\end{quote}

Hours of interviews and observations, pages of transcripts and field notes, and the large number of codes, labels, and post-it notes that follow are a familiar sight to any qualitative researcher. Qualitative research, known for its ability to deeply understand rich in-situ practices and experiences, is widely used in CSCW and HCI for developing human-centered understandings of computing systems. However, qualitative research is also, by its nature, manual and laborious. While the scale of data generated by methods like interviews and ethnography can be viewed as ``small'' in our current world of ``big data,'' it nevertheless can be overwhelming to humans, and requires tremendous human effort to analyze. Other qualitative methods like content analysis, on the other hand, often directly tackle the realm of ``big data'' by using samples like millions of scraped online posts (e.g., \cite{brubaker_we_2011}), a scale even more clearly impossible for humans to handle. Therefore, qualitative researchers use methods such as subsampling \cite{robinson_sampling_2014} to reduce the sample size to something humans are capable of handling, forgoing the opportunity to take advantage of the richness offered by the full dataset.

The challenge of qualitatively analyzing big data has led some to consider the potential for artificial intelligence (AI) to assist qualitative researchers. While one may argue that qualitative research and AI are incompatible due to their different epistemological traditions, recent work has discussed the conceptual similarities between machine learning (ML), a technique often used in AI, and grounded theory methods (GTM), and proposed hybrid models that take advantage of both types of methodology \cite{muller_machine_2016}. Putting this comparison to the test, one study has even shown empirically that GTM and ML can produce similar results on the same data \cite{baumer_comparing_2017}. There also have been various efforts to integrate ML into qualitative research (e.g., \cite{yan_optimizing_2014, marathe_semi-automated_2018}), aiming to automate deductive steps of the qualitative research process---applying established codebooks to the dataset---and reduce human effort. However, the inductive step of generating the codebook through discovering patterns and themes cannot be overlooked. Is it possible that AI can also support the inductive step of qualitative research that finds meaning in unstructured data instead of applying already-found meaning on a larger set of data? While human-AI collaboration seems promising for qualitative research, is it something that qualitative researchers themselves desire? And what abilities and qualities do people seek (or not) in AI that would make it a good collaborator? Answers to this question will not only shed light on how we can better design AI assistance for qualitative researchers, but also provide deeper insights into human-AI collaboration in fields typically seen as human-only. n this paper, we do not seek to examine the efficacy or promise of particular models or algorithms, but to uncover the characteristics and needs unique to qualitative research to which AI designers and developers should pay attention. While qualitative research includes many subprocesses such as data collection and result reporting in additional to data analysis, in this paper, we focus on the analysis process in particular because it shows the most promise for potential AI assistance, as both prior research and our participants suggest.

In this work, we examine the practices of qualitative researchers, and the capabilities that they do (or do not) want assistive systems to have, through an analysis of 17 semi-structured interviews with qualitative researchers. While the participants did not participate in the interviews with AI or machine learning in mind, participants imagined systems that were able to automatically offer intelligent inferences and suggestions when they were asked what their imaginary ``perfect tool'' would be. In this paper, we collectively refer to these systems as ``AI'', following the rationale in prior human-AI collaboration literature in CSCW \cite{cai_hello_2019}. We first describe the research methods that they use in their work, and the kinds of assistance they would like. Next, we describe the collaboration practices of participants whose methods involve collaboration, highlighting that reaching consensus is a nuanced social practice that extends beyond the actual research. We then discuss participants' love-hate relationship with qualitative analysis: While the vagueness of qualitative analysis induces confusion and self-doubt, this doubt is essential to their research practice and AI should not remove that doubt in their analysis. Using a framework of task delegability, we close by discussing how qualitative research is a unique case that deviates from the typical tasks humans delegate to AI. While we do not (nor can we) prescribe exactly what algorithms or models should be used, we argue that honoring uncertainty and serendipity should be a central quality of AI that assists with qualitative research, where data labels are fluid and ``negatives'' are not well-defined or sometimes nonexistent.

\section{Related Work}

\subsection{Qualitative Research and Methods}

Qualitative research accounts for a substantial proportion of CSCW and social computing scholarship---more than half of the CSCW 2019 papers use qualitative methods \cite{gilbert_cscw_2019}, with new opportunities for qualitative research being a constant topic in CSCW as well \cite{fiesler_qualitative_2019}. Qualitative researchers use a number of different analytic methods to generate insights, leveraging both inductive logic (to build patterns and themes from the bottom up by organizing data into increasingly abstract units of information) and deductive logic (to check existing patterns and theories against data in a top-down fashion). For example, thematic analysis \cite{braun_using_2006}  and constant comparative method \cite{glaser_constant_1965} both describe processes of identifying, analyzing, and reporting emergent patterns within a set of data. Informed by the perspective of hermeneutics, interpretive textual analysis, a commonly used method in qualitative research, aims to understand cultures and subcultures through situated interpretation of texts \cite{allen_textual_2017}. In addition to these methods that focus on analysis, grounded theory \cite{glaser_discovery_1967, strauss_grounded_1997} alternates collecting and analyzing qualitative data in order to generate theories that are ``grounded'' in that data \cite{charmaz_constructing_2006}. It provides a set of systematic guidelines for analysis while maintaining flexibility to fit the given set of data. Data sources in qualitative research are also diverse and overlapping, and any of the five traditions above could draw from interview data, field notes, and textual or visual content \cite{creswell_qualitative_2013}. Many qualitative approaches make use of open qualitative coding \cite{strauss_grounded_1997} to inductively label and categorize emergent concepts while maintaining theoretical freedom, which is particularly useful for sifting through large amounts of unstructured data. Such inductively generated coding schemas can then be deductively applied on the full set of data.

Strengths of qualitative work include the ability to examine a space in great detail and depth, the adaptability of the research framework, collection of data with subtleties and complexities not available in quantitative data, and the situating of data in the context of experience and community \cite{ryan_introduction_2018}. Compared to quantitative research that often relies on positivism that believes that there are observable, measurable, ``objective'' facts that hold true for everyone, qualitative research often favors interpretivism, which argues that knowledge and truth are subjective and situated, and dependent on people's experiences and understandings \cite{ryan_introduction_2018}. Instead of goals such as causal determination, prediction, and generalization sought by quantitative researchers, the goals of qualitative inquiries are ``illumination, understanding, and extrapolation to similar situations'' \cite{golafshani_understanding_2003}. 

However, there are two major challenges of qualitative work. First, qualitative analysis is largely manual and laborious. Qualitative researchers have long used ``cut and paste'' methods for separating chunks of information, either physically (e.g., index cards) or in databases or spreadsheets \cite{campbell_technology_1997}. There are now a number of qualitative analysis software tools (e.g., MaxQDA, ATLAS.ti, Dedoose) in which researchers can highlight and assign labels to portions of text, and these tools can produce some lightweight statistics (e.g., count) of these labels, with potentials of having more advanced features like topic modeling. While computational tools can relieve qualitative researchers of the burden of data management, their effects on data analysis are still highly debatable. On the one hand, qualitative researchers have argued that increasing the sample size in qualitative research designs is not an advantage, and researchers should focus on as many distinct cases as possible rather than as many cases as possible  \cite{wiedemann_opening_2013}. Using software tools that generate coding schemes based on a large number of documents risks opportunities for creativity and serendipity that qualitative research is uniquely able to provide \cite{wiedemann_opening_2013}. On the other hand, current qualitative methods such as content analysis \cite{krippendorff_content_2004} can require researchers to handle such an overwhelming amount of data that researchers may need to rely on subsampling or strategically focusing on a subset of data \cite{robinson_sampling_2014}. However, it can be difficult, if not impossible, to extrapolate the findings from the analysis on a subset of the data to the original dataset.

Second, the rigor of qualitative analysis is more difficult to assess and demonstrate than research rooted in positivist traditions that relies on measurement. There has long been a wealth of literature from multiple disciplines focusing on the assessment of the reliability and validity of qualitative analysis, providing guidance for publication \cite{elliott_evolving_1999}, procedures for developing coding schema  \cite{campbell_coding_2013}, and design research approaches to ensure rigor in qualitative analysis \cite{maher_ensuring_2018}. CSCW and HCI scholarship have also taken a keen interest in meta-studies of qualitative methods, such as comparing interview methodology through different media \cite{dimond_qualitative_2012} and characterizing localized standards of sample size at CHI \cite{caine_local_2016}. McDonald et al. \cite{mcdonald_reliability_2019} have shown that some CSCW and CHI papers chose to report interrater reliability (IRR) even if IRR might not be compatible with their research design, which highlights a need for demonstrating reliability in qualitative research. These two challenges suggest that qualitative research can benefit from new approaches that aim to help with handling larger amounts of data, and our study explores AI as an option to support qualitative scholars with their research.

\subsection{Machine Learning and Qualitative Research}

Machine learning (ML) techniques are known for their ability to extract patterns from small samples and generalize them to larger datasets, show promising opportunities to apply traditionally ``small'' data-focused qualitative methods to ``big'' data. Muller et al. \cite{muller_machine_2016} discussed conceptual similarities between machine learning and qualitative grounded theory methods (GTM)---they both make major claims to be grounded in the data, and they both start with and return to the data. Through a discussion of how Straussian GTM and Glaserian GTM (the two major approaches to GTM) are similar to unsupervised and supervised learning respectively, Muller et al. proposed a number of hybrid iterative approaches that integrate machine learning and GTM. Baumer et al. \cite{baumer_comparing_2017}, through an empirical study, showed that ML and GTM could produce strikingly similar output from analyzing the same survey data, and later further showcased the promise of ML through their Topicalizer system \cite{baumer_topicalizer_2020}. They argue the promise of combining ML and GTM, but also caution that the results generated from machine learning models should be a scaffold for human interpretation instead of definitive answers, a point we echo in our analysis. 

Another thread of research has explored the efficacy of using machine learning as a tool to support qualitative research, with an intention to automate parts of the research process. Yan et al. \cite{yan_optimizing_2014} experimented with a human-in-the-loop approach by having a machine learning model code the entire dataset from a pre-defined codebook and having human annotators later correct the labeling, and concluded that creating a one-size-fits-all ML model for all codes in a multi-dimensional coding scheme was impossible. Paredes et al.'s \cite{paredes_inquire_2017} Inquire tool aimed to help qualitative researchers by uncovering semantically-similar content in large-scale social media texts. Marathe and Toyama \cite{marathe_semi-automated_2018} explored whether qualitative coding could be partially automated through a formative user study and user testing of a prototype. They found that repetitive coding with a well-developed codebook lends itself nicely to automation, but that having good IRR is crucial for automatability. Chen et al. \cite{chen_using_2018} pointed out that the goal of performance optimization in machine learning may be at odds with the goal of discovering patterns in qualitative research where the categories evolve over time, and suggested that a model that identifies points of ambiguity (i.e., disagreements in coding) may be more useful to qualitative researchers. It is important to note that the studies above focused on deductive aspects of qualitative research, typically with well-defined codebooks, and not inductive practices.  Our exploration extends this prior research by providing insights into the inductive step of the initial discovery of patterns and themes.

\subsection{Human-AI Collaboration}
The research we have surveyed thus far shows the infeasibility of automating qualitative analysis entirely, but also points towards ways that AI could be a collaborator that works alongside humans, rather than a delegate that performs specific tasks. Human-AI collaboration itself is increasingly gaining importance in CSCW scholarship. Emerging work focuses on a variety of topics from information extraction \cite{mackeprang_discovering_2019} to image understanding \cite{zhang_dissonance_2019}, consistently highlighting the need for transparency and explainability of AI. Mackeprang et al. specifically pointed out that at high levels of automation, users were unlikely to challenge results provided by AI when unaccompanied by explanations \cite{mackeprang_discovering_2019}. This situation, to say the least, would be undesirable in the context of qualitative research. Based on interviews with data scientists about their perception of collaborating with AI, Wang et al. reiterated the importance of AI transparency, and further noted that AI should be designed to augment, rather than to automate, a point we echo in our findings \cite{wang_human-ai_2019}. Furthering this concept of augmentation, Yang et al. proposed the idea of ``unremarkable AI,'' whose interaction with humans should have the right level of unremarkableness but should significantly improve their work \cite{yang_unremarkable_2019}. Although Yang et al.'s study is situated in the context of high-stake clinical decision making processes, we see the notion of unremarkable AI as applicable to other contexts where AI should not be obtrusive. 

For human-AI collaborations to be successful, the issue of what work the AI should or should not perform needs to be addressed. Lubars and Tan proposed a framework for task delegability to AI \cite{lubars_ask_2019}. They enumerate four major factors to consider: the human's motivation in undertaking the task, difficulty of the task, risk associated with incorrectly completing the task, and trust in the AI's ability to accomplish the human's goal. These factors, especially risk and trust, resonate with existing concerns around AI fairness, accountability, and transparency that have been increasingly present in human-AI collaboration scholarship. In this study, we use this task delegability framework to look at the case of qualitative analysis, and show how qualitative analysis is a uniquely challenging case for human-AI collaboration.

\section{Methods}
We conducted interviews to investigate the qualitative research methods used across various fields, with specific focus given to the coding process and tools that assist in data analysis. After acquiring IRB approval from our institution, we focused recruitment efforts on participants who primarily use qualitative research methods in their work. We recruited participants via emails to mailing lists of departments that regularly conduct qualitative research (e.g., communication, anthropology) at our institution, as well as public posts on social media that invited CSCW 2019 attendees to participate. We also encourage participants to share the call for participation with other people, resulting in a snowball sample. 

The first two authors conducted 17 semi-structured interviews with qualitative researchers at our institution and among the broader CSCW qualitative research community. Participants were majority women with 2 men, 1 non-binary, and 1 agender individual.  Participants ranged in age from 25 to 45, from the fields of Information Science, Anthropology, HCI, Linguistics, and Journalism. A majority of the participants were from the US, in addition to one from the UK, one from Japan, and one from Switzerland. We do not map this specific demographic information to individual participants in order to ensure that they are not individually identifiable. Fourteen interviews were conducted in-person, and three remote. Interviews lasted from 20 to 70 minutes, depending on the depth of the responses given. Table \ref{tab:demographics} lists our participants along with their field of study and academic position.  Participants were compensated with a \$30 Amazon gift card upon completion of the interview.

\input{participants-table.tex}

Our interview protocol focused on how qualitative researchers collect and analyze their data, taking a broad exploratory approach. Interview questions included what methods the participants use, if there are any tools they use to assist in their work, attitudes and opinions about qualitative coding as a process, and relationship dynamics between collaborators. During the interviews, we asked participants to describe a current or most recent qualitative research project they have conducted, including scope, scale, and timing, in as much detail as possible. We then asked participants to describe their collaboration practices if relevant. The next section of questions focused on the software and tools the participants use for their data collection and analysis. We also asked participants for their attitudes and opinions on the coding process, including what they enjoyed about it and where there might be pain points. Finally, we asked participants to talk about what kind of assistance they would like if they had a perfect tool, and areas of their research where they did not want assistance. As mentioned in the introduction, we did not specifically ask participants about ``AI assistance'' so as to prevent leading them to answers about the term ``AI'' without a shared, concrete idea of what ``AI'' means. Instead, we asked what kind of assistance that they envisioned in a perfect tool, and their responses led us to consider AI as a potential solution, and ultimately human-AI collaboration as the focus of this paper.

We performed a thematic analysis of the interview transcripts \cite{braun_using_2006}. Prior to analysis, all interviews were transcribed, anonymized, and assigned the participant IDs presented here. The first author initially engaged in one round of open coding, using the software MaxQDA. Then the second author, after reading the transcripts, discussed preliminary emerging code groups such as ``help with the mess'' or ``feeling of doing it right'' with the first author. Two more rounds of iterative coding helped us combine similar code groups into higher order categories such as ``disciplinary differences.'' The first author used these categories to produce a set of descriptive theme memos \cite{saldana_coding_2009} that described each category with grounding in the interview data. All authors then discussed and updated the memos regularly to reveal the relationships between the categories and finally clarified the themes, which resulted in the five main findings we discuss below.

\section{Findings}

Our findings document qualitative researchers' current practices and needs that AI needs to support, and discuss their unique, nuanced challenges where AI should not be applied. We begin by describing the methods participants used in their qualitative research, and the kinds of assistance that participants desired from software tools. Next, we discuss the complex collaboration dynamics in collaboration-heavy disciplines by highlighting that reaching consensus is a nuanced social practice that extends beyond the discussion of the research project itself. We then discuss participants' struggle with confidently conducting qualitative research, highlighting the central role of uncertainty in qualitative analysis. Finally, we discuss participants' connection with and need for agency over the analysis of their data despite their self-doubt. From here, we conclude by revealing the bottom line that AI assistance should not cross---they should not replace human researchers in doing the analysis.

\subsection{Qualitative Research Practices}

\input{methods-table-1.tex}
\input{methods-table-2}

Our participants used a variety of methods in their qualitative research. Among them, interviews were the most common data collection method, used by 15 participants across 6 fields. Others that were common included participant observation (7 participants), and note taking (4 participants). Participants also described data collection methods specific to their field or discipline, such as Jeffersonian transcribing in linguistics, oral history in journalism, and ethnography in anthropology. Compared to the varied data collection methods, participants' descriptions of their analysis methods, while having varied names and granular details, broadly converged to a process of inductively identifying interesting pieces of data then finding patterns among them. While we acknowledge that qualitative methods entail a much broader range than what our participants reported, our analysis here only focuses on the methods our participants used. Table \ref{tab:methods} shows the complete list of research methods that participants used. 

Surveying the practices of qualitative researchers runs into some challenges due to terminology. While many of the participants share similar research practices, they referred to these practices with different terms. These differences not only exist across disciplines, but within disciplines as well. For example, P06 referred to the process of inductively building a codebook then deductively labeling according to the codebook as ``thematic analysis,'' while P01 and P12 called the same process ``content analysis.'' Many participants referred to the common grounded theory research method of first identifying data snippets of interest and assigning them descriptive labels, then generating deeper insights from these labels (sometimes with the explicit process of categorizing these labels). However, participants have different terms for both steps. Terms for the first step---identifying interesting units of data---included ``qualitative coding,'' ``open coding,'' ``inductive coding,'' and ``highlighting unusual things.'' Participants’ terms for insight generation are even more varied, from ``memoing'' ``affinity mapping'' to descriptive phrases without a proper term---``seeing themes,'' ``grouping into categories,'' and ``classification process to organize data into themes.'' Furthermore, while P04, a Ph.D. student in linguistic anthropology, also used the term ``coding,'' she was referring to transcribing the interviews into the International Phonetic Alphabet, instead of identifying interesting data snippets---something we only realized halfway into our interview with her. 

Overall, we saw participants use inconsistent vocabularies despite referring to similar methods. While we have not heard stories of participants running into communication problems, it is possible for researchers to accidentally miscommunicate research methods and for reviewers to have confusing expectations. Assistive tools can also create barriers by choosing to use language that favors one particular genre of academic training.

\subsection{Assistance From Computational Tools}
Qualitative research is a manual, labor-intensive process that would benefit greatly from computational tools, but for many participants, existing tools were not sufficient for their needs. In this section, we first describe the gaps in current tools, and then discuss the assistance that participants desired.

\subsubsection{The Inadequacy of Existing Tools.}

Participants used a variety of software to assist with qualitative research, and only three participants did not use any such tools (see Table \ref{tab:methods}). However, not only did many tools fail participants in helping them handle large amounts of qualitative data, participants were also unwilling to learn more sophisticated tools. According to P05, the amount of notes and labels quickly became overwhelming, and the software tool that she used was insufficient:

\begin{quote}
    I do everything in Google Docs. You've got 500 codes first and then you've got notes everywhere all over them, which I'm currently trying to decipher and I'm kicking myself for not having a better method. (P05)
\end{quote}

Many participants used word processing software like Google Docs to code qualitative data, but these software were not able to handle the overwhelming amount of codes and notes, as was the case for P05. While some participants were aware of other software tools that have more sophisticated functionalities, they did not use or stopped using them for various reasons. For example, P10 stopped using ATLAS.ti because the software broke frequently on her computer, so she returned to her old method of using Excel and handwritten notes. P09 only used software to organize his transcripts, and told us that learning how to use more complicated software was not worth the effort:

\begin{quote}
    Using any tools, I think it gets in the way of the analysis...I think the focus then inevitably becomes on the tool and how I can manipulate and push data in order to make it appropriate for the tool... I need to figure out how I can fluster data in a certain tool, in a certain way so as for it to make it easier for me later on. And I was just thinking it's a waste of time. (P09)
\end{quote}

P09's quote is exemplary of participants who chose not to use software tools. Their anticipated cost was not only in learning how to work the software, but also in how to use the software in their own way that is sustainable in the future, and these two layers of barriers precluded these participants from using it. P14, while kept using the qualitative analysis software NVivo, limited herself to the most basic functionalities of coding data and making code categories, then hand-copied these categories onto post-it notes that she later put on the wall. Therefore, participants wanted ``more intuitive ways to do qualitative analysis,'' and something that could help them ``sort through this mess'':

\begin{quote}
    If I had a [perfect tool] I would like a little bit more input in putting together connections between the data. (P12)
\end{quote}

\subsubsection{Desired Assistance.}
Participants had different ideas of how a tool could help them make sense of the connections within the data, and cross-sectional analysis and visualization features were common requests:
\begin{quote}
    I really like when I can slice and dice the data. So I want to be able to say, okay, all the people I've talked to who identify as queer, how did they feel about capitalism? I want to be able to do a cross-sectional analysis on multiple codes and domains. (P12)
\end{quote}

\begin{quote}
    Just better visualizing the relationships in the data itself. So if I'm missing out codes because I have less time, or if my co-author isn't being participative enough or whatever, then I want to be assured of the fact that I was able to see all kinds of relationships. (P07)
\end{quote}

It is important to note that cross-sectional analysis and visualization features, though implemented to various extents, already widely exist in qualitative analysis software such as MaxQDA and NVivo. Prior research has noted that qualitative researchers usually confine themselves to the most basic features of qualitative analysis software \cite{wiedemann_opening_2013}, but here we see that the participants still want these seemingly basic features, which suggests that the participants never discovered these features in the first place due to the overall difficulty of using the software. Furthermore, P07 specifically pointed out that the tool should not be ``like the NVivo type, where I have to really learn a lot of it.'' In other words, P07 admitted that an existing tool may be able to achieve what she wanted, but the bar of learning the software was too high. 

While participants across disciplines wanted additional help to make sense of the data, some participants would appreciate features that are specific to their fields. For example, P05, an anthropology Ph.D. student, requested features designed for their own research tradition:
\begin{quote}
    Maybe, I don't want to say a translation tool, but kind of almost a sort of node system to be like ... ``These were the literal words that were used, but this is kind of the underlying meaning to that.'' (P05)
\end{quote}

While anthropologists might benefit from features to help them note underlying meanings to spoken words, P15, a Linguistics student, told us that being able to get down to the most granular details of speech was important:

\begin{quote}
    I'm thinking especially in my field where when I'm transcribing interviews, I'm actually not just looking at what is literally being said. There's a process of transcribing for tone of voice, gestures, and things. (P15)
\end{quote}

Unlike many of our other participants, for whom the purpose of transcribing is to transform audio data into textual data that is readily analyzable, the detailed transcription---called Jeffersonian transcription---\emph{is} the core of P15's analysis. Her analysis did not include categorizing snippets of labeled data. Her unique genre of analysis means that a tool designed only for the most common use case will be of no use to her at all.

In addition to features that handle the mechanics and the presentation of qualitative data, participants also wanted features that could take one step further by providing suggestions and inferences, based on the way that the researcher analyzes the data:

\begin{quote}
    If there was some sort of learning algorithm, for example, that would suggest ... other quotes that were similar to that one. (P08)
\end{quote}

\begin{quote}
    Or maybe even just the ability to be based on whatever magical software anyone can develop being, ``based on the fact that you tagged it this way, we think that these are the more important interviews or not important, but we think that maybe you want to pay more attention to these.'' Or be like, ``You completely neglected this one. Did you forget about it or is there actually nothing here for you?'' (P05)
\end{quote}

In sum, participants wanted various types of help from computational tools but often refrained from using sophisticated tools due to their steep learning curve. While their reluctance may indicate usability issues with such tools, their reluctance also suggests opportunities where AI assistance could provide them the desired data representation and visualization, and only offer more sophisticated explanations and customizations if desired. The promise of AI assistance is further highlighted by participants' imagined intelligent suggestions---or, in P05's words, ``magical software.''

\subsection{Collaboration Dynamics}
Qualitative research can often be collaborative, involving multiple researchers examining and then reaching consensus on the same data to increase reliability and reduce individual biases. Participants from Information Science and HCI told us that collaborative work with co-authors was common, so it is no surprise that, when asked about features they would like in their imaginary perfect tool, only Information Science and HCI participants mentioned collaboration features:

\begin{quote}
    While I was maintaining this online notebook, I could hypothetically, if my co-author had enough time, the dream scenario would be that he or she would be reading those notes every day and commenting on them ... So the collaborator element's definitely important and could help a lot with the eventual coding and analysis process, even while you collect data. (P07)
\end{quote}

With collaboration being the norm, a natural follow-up question is how to reach consensus when there are multiple researchers on a single project. ``Consensus,'' however, takes on many different flavors and nuances, according to the participants. 

Inter-rater reliability (IRR) is often used to calculate consensus during a deductive phase of coding, but is often inappropriate for inductive analysis. Consistent with McDonald et al. 's findings \cite{mcdonald_reliability_2019}, reaching consensus through IRR was rare for our participants---only two used IRR to verify consensus, and two others mentioned IRR only to note that they did \emph{not} use it in their research, likely because our participants largely conducted interviews, of which the analyisis is often incompatible with IRR. 

Our participants most commonly perceived consensus as some flavor of ``everyone agrees,'' though agreement sometimes can be a flexible and loose standard:

\begin{quote}
    Once everyone is pretty much satisfied and doesn't have any complaints or questions or comments anymore that are significant, and by significant, I mean like ``we have to change this or it's going to ruin the study.'' (P02)
\end{quote}

Compared to P02, P14's standard for consensus was even more flexible---there is no definition for consensus; it is something that the researcher understands intuitively as they are doing the research:

\begin{quote}
    [Consensus is] just like a kind of tacit knowledge thing where it's you know when you see it, but it's very hard to actually kind of define. (P14)
\end{quote}

P14's view of consensus as fluid highlights how it is the outcome of nuanced social practices that is not easily explainable or definable. However, its flexible nature also means that achieving consensus might depend on the researchers' established, implicit norms generated from having worked together for an extended period of time, a facet often overlooked by prior conceptualizations of researcher consensus (e.g., \cite{macqueen_codebook_1998}).

These social practices, as we learned from participants, often reduce ``consensus'' to the decision of a person with the ultimate decision-making power, rather than multiple researchers having reached some level of agreement. The decision maker, however, varies. Sometimes the decision maker is the project leader:

\begin{quote}
    If I'm leading the paper, to me it's my decision [if we have reached consensus]. (P08)
\end{quote}

\begin{quote}
    If the stakes are high, if I am the primary investor in a particular project then yes, I think I should take the decision on this. But if it's maybe my advisor's project I think I would give her the final say in that. (P10)
\end{quote}

Sometimes the person with the final say is the most senior person, who may or may not be the project leader:

\begin{quote}
    If [my collaborators] don't [agree], we have a little bit of a debate and I'm assuming the most senior person probably wins. (P03) 
\end{quote}

While P03 did not personally have experience with the most senior person ``winning'' the debate, as a junior Ph.D. student this was their default assumption, which suggests a tendency for researchers (especially junior ones) to acquiesce to authoritative figures. P07 elaborated more on this tendency, pointing out that ``consensus'' is the result of an negotiation of collaboration dynamic and power:

\begin{quote}
   I think it depends a lot on who the collaborators are and what the power relationships with those collaborators are ... These things are not objective ... It's not just about what codes appear or what themes appear the most salient and relevant. So I think it's more about the collaboration dynamics, the power, all of those things, rather than what the codes are seeing ... It's literally about [power]. (P07) 
\end{quote}

While it might seem ideal for every researcher in a team to have their opinion treated equally, P07 directly refuted this notion, claiming that it would only jeopardize the project:

\begin{quote}
    [Collaborative qualitative analysis] would still definitely take on some hierarchy in what I imagine. There would still be someone who's written more papers. There will still be someone who's written fewer papers. Or still have an expert on the topic, versus someone who's not the expert on the topic. ... I can't imagine there being a flat hierarchy somewhere. And actually, that one time that it happened ... things went really down south. The paper was about everything. It got rejected. Then there was a fight. Then there were all these conversations about who's contributed the most, who's going to be the first author and so on. (P07)
\end{quote}

MacQueen et al. \cite{macqueen_codebook_1998} argued that in team-based qualitative analysis it is good practice to have one person maintain the team codebook. While MacQueen et al.'s recommendation largely came from a project management standpoint and maintained that the codebook development should still be done collaboratively, here we see a case where one person being in charge of the \emph{direction of analysis} can be more beneficial than a true, equally collaborative fashion. Overall, we observe that in fields where collaboration is common, reaching consensus among researchers can often be a nuanced social practice beyond mere agreement on the research product.

\subsection{The Existential Crisis of Qualitative Researchers}
Despite the variety in discipline-specific research methods, desired help from software tools, and meanings of consensus, participants across disciplines consistently told us about one feeling: qualitative research is hard and messy.

\begin{quote}
    Sometimes it can be a little bit frustrating when you're sort of in that process where you're like, ``Is this a theme? Is that a theme?'' Then sometimes when you're going through ... a couple of interviews in and you're like, "Oh there's a theme," and then you're like, ``Oh fuck, that's probably a theme'' ... so then you have to go back to see if that [theme] is in [the other interviews]. (P13)
\end{quote}

Like P13, many participants described how qualitative analysis could be frustrating, and this frustration often comes from researchers' self-doubt:

\begin{quote}
I'll just code, and maybe it'll be relevant and maybe it won't be, and I'm not really sure ... If someone just mentions it in passing, do I use that? ... If someone has a few sentences about it then do I code that? Like what is useful to actually code? (P03)
\end{quote}

\begin{quote}
    Am I drawing connections between things that aren't necessarily connected? ... which also goes back to not having time to [follow up with] people and be like, ``Hey, so you said this thing, I'm just wondering, can you expand that more?'' (P05)
\end{quote}

P05 doubted herself when drawing what she feared were non-existent connections in the data, and attributed it to having conducted imperfect interviews. The guilt of imperfect research was common among participants, who felt that they could have asked better questions during interviews, could have used more data in the final research product, or could have done a deeper analysis if they had had more time:

\begin{quote}
    I didn't really get to do maybe as much coding as I would've liked, or to look at as many complex issues as I would've liked. ... It was a restriction on time. (P03)
\end{quote}

In addition to imperfect research execution being a common source of guilt, others felt apologetic for not following an ideal research methodological paradigm:

\begin{quote}
    I think I would be lying if I said that I completely allow myself to be driven by the data ... especially this study, for example. I did it to inform the design of potential technology that I wanted to build, while I'm doing the questions, I was already looking for particular things. ... So I think that already in the questions you have implicit codes. (P11)
\end{quote}

\begin{quote}
    I feel like we're doing it wrong. ... Especially because [we are] saying these are the topics that we're interested in. That's not very grounded theory of me. But, I don't quite know how to do it otherwise, because obviously ... I'm going to look at this interview already from a different perspective than, say, a psychologist. (P15)
\end{quote}

Participants thought their own research execution was imperfect---for example, ``not very grounded theory.'' However, in participants' own unique research projects, the ``perfect'' methodology may not be the most desirable approach. While HCI (as well as CSCW)  has distinguished itself as a ``discipline that has often proceeded with something of a mix-and-match approach'' \cite{dourish_reading_2014}, here we see evidence that such a hybrid paradigm---for example, being driven by the data but also having a topical focus---is also practiced across disciplines.

Overall, self-doubt with all parts of the research process was pervasive among participants. This self-doubt was so strong that P07 told us the moment she had to approach the data she collected was what she ``dread[ed] the most.'' Participants' shared experience of self-doubt highlights the inductive and interpretive nature of qualitative research, and suggests that this vagueness may be inevitable and essential to the qualitative research process itself. While it might be tempting to design AI that can reduce qualitative scholars' anxiety by providing concrete suggestions from their data, we found ample reasons why researchers might resist such assistance, which we detail in the next section.

\subsection{Don't Do It for Me}
Even though participants consistently told us how qualitative research can be full of doubt and vagueness, they did not want AI assistance to eliminate that vagueness if it meant eliminating the work to make sense of their messy data. Participants told us that despite the uncertainties, the new discoveries they made through their research process were the most rewarding: 

\begin{quote}
    It's really satisfying. ... It's those kinds of exciting Eureka moments that make research kind of worth it. (P14)
\end{quote}

\begin{quote}
    My favorite style of coding is more where I'm able to bring out all of these different things. ... So that's what I really enjoy, the process of recollecting everything that happened, and then coming up with something. (P07)
\end{quote}

Participants' comments on their unexpected new findings show that serendipity was a major reason why participants enjoyed qualitative research. Integrity was a major factor as well. P12, for example, described  a sense of responsibility toward her project:

\begin{quote}
    How do I feel when I'm coding? Overwhelmed. I feel such a great sense of stewardship in the project that we've been discussing. I want to get it right and that makes me feel a lot of pressure. (P12)
\end{quote}

The motivations and responsibilities that researchers feel around their data and analysis may present some hurdles for any tool meant to support their work. While some participants mentioned that they would benefit from AI-generated suggestions and inferences, these participants, with their sense of serendipity, responsibility, and intimacy toward their research, explicitly expressed that they needed agency over the analysis of their data, and AI should not do it for them.

\begin{quote}
    I don't want anything to do the analysis for me. That doesn't make sense. This is my interpretation of things ... It's one of the most intimate things. ... I have these little like research crushes with some codes or participants. ... These [data] are my babies. (P11)
\end{quote}

P16, while being in Linguistics and doing a completely different kind of analysis than P11's, shared P11's feelings:

\begin{quote}
    As nice as it theoretically sounds to have something do automatic transcription for you, I actually would not want that even if it could do the Jeffersonian level of detail because it's so important for you as a researcher to understand what's happening in your data, doing it yourself. ... Your transcription is part of your analysis as well and that's the only way for you to get your hands dirty with the data is to get in and see what's happening with it. ... I don't know of anybody who uses [professional transcription services] because you take a lot of pride and ownership over your data and it's important to your participants and I would never want to farm that out when it's your research in your research process. (P16)
\end{quote}

To P11 and P16, outsourcing analysis for them not only impacts research quality, it also robs them of their emotional connection---the intimacy, pride, and ownership---with the data. Compared to P11 and P16, P06 was slightly more open to the idea of AI-assisted analysis. However, P06 also pointed out that the tool should be limited to the level of suggestions and that making decisions would cross a line: 

\begin{quote}
    I don't want it to take out extra information automatically. I don't want it to try to guess or try to ... well, it can try to guess, but I don't want it to decide. ... It can do a suggestion, but I'm not sure if I would trust the suggestion, or actually it can suggest the discussion points that we need, and then we can discuss rather than decide it for us. (P06)
\end{quote}

In contrast, P14 felt even suggestions could be harmful to her research process: 

\begin{quote}
    Maybe it could make suggestions, but even then I don't know if I want it because it doesn't know what my research questions are. ... I don't want it to pick the good quotes for me, that's for sure. ... I don't want the computer to do it. I also don't want the computer to influence how I'm thinking about it. ... I don't necessarily want to be primed or how that's going to affect my thinking ... if it's making this suggestion and now I can't unsee it. (P14)
\end{quote}

While participants generally resisted complete automation of qualitative analysis, P06 and P14's comments show that even a low-level, suggestion-based automation (according to Parasuraman et al.’s model of autonomy \cite{parasuraman_model_2000}) can also be undesirable. During the interview, P14 asked if there was a machine to interfere, then ``why bother?''---why bother having a human researcher at all? According to P14, even if the researcher does not necessarily have to use or trust the suggestions, seeing the suggestions alone can negatively impact the researcher's interpretation and analysis, or even worse, eliminate the meaning of human researchers.

Overall, participants told us about their conflicted feelings about qualitative analysis, one that is marked by anxiety but also intimacy and ownership. Driven by these feelings, they resisted AI to replace them in their analysis, despite seeing its potential to augment their work. These findings suggest that, unlike tasks typical for AI to accomplish, the objective of qualitative research is, often intentionally, ambiguous. Furthermore, qualitative researchers value this ambiguity because it offers them serendipity and opportunities for deep insights. In the discussion section, we detail how qualitative research is a unique case for human-AI collaboration, and provide design implications for AI assistance to honor ambiguity and serendipity.

\section{Human-AI Collaboration in Qualitative Research}
Our findings reveal complicated feelings participants had toward the practices involved in qualitative analysis. On the one hand, participants struggled with self-doubt in their research analysis and saw the appeal of AI assistance to help them see relationships behind their data. On the other hand, they also liked the analysis process as it is and felt connected with their data, therefore skeptical of AI driving the analysis. This tension between the desire of AI assistance and the need for human agency suggests opportunities for human-AI collaboration in qualitative research, but also draws a line that AI assistance should not cross. 

To unpack this tension, we turn to a framework proposed by Lubars and Tan to describe the delegability of a task to AI \cite{lubars_ask_2019}. While there were likely many small steps in the qualitative research process (e.g., open coding, axial coding, synthesizing), our participants' resistance of complete automation suggested that they viewed qualitative analysis as a singular task in the context of AI assistance, and we can reasonably speculate that AI designers (who are arguably less familiar with qualitative analysis) might do the same. In their framework, they enumerate four factors that must be considered when determining the appropriateness of delegating a human task to AI: \textbf{motivation}, \textbf{difficulty}, \textbf{risk}, and \textbf{trust}. 

Lubars and Tan conceptualized motivation as having three components: the human's intrinsic motivation of carrying out the task, whether the human's goal itself is mastering the task, and the utility or importance of the task to the human. Our participants' enjoyment and pride in conducting qualitative research, as well as their wish to improve their research skill, show that qualitative research involves high motivation: participants enjoyed and were committed to conducting their analyses. While one may assume that high motivation indicates low delegability, Lubars and Tan did not find statistically significant correlation between the two variables. The lack of correlation coincides with the sentiments of our participants, who were invested in their qualitative work but also acknowledged the appeal of appropriate AI assistance.

Difficulty also consists of three components: a person's perceived ability to complete the task, as well as the required human effort and expertise. Lubars and Tan found a general negative correlation between difficulty and task delegability: the more able a person perceives themselves to complete the task, the more human effort and expertise required by the task, the less delegable the task is to AI. Qualitative research can be considered high difficulty, as it requires effort and expertise on behalf of the researcher, which suggests its low delegability to AI according to Lubars' and Tan's findings. The difficulty of analysis is evidenced by the low confidence participants had in their analysis. While Lubars and Tan also use required social skills and creativity as dimensions to contextualize the previous three dimensions, they are not central to the conceptualization of difficulty, though one can reasonably argue that qualitative research requires high levels of both social skills and creativity. 

Additionally, in the survey that Lubars and Tan used to evaluate their framework, they observed that self-confidence and intrinsic motivations were positively correlated---people enjoy doing tasks that they are good at. However, the case of qualitative research serves as a unique counterexample: not only do people enjoy doing something they are not confident about whether they are doing it correctly, but low-confidence itself is central to the outcome of qualitative research. Therefore, our study on qualitative research challenges the current task delegability framework, and also raises questions around the integration of AI into qualitative research: Should we consider qualitative research as a single, unified ``task'' for AI to assist with? If qualitative research as a single task should not be delegated, what parts of it are appropriate for AI to assist with?

Qualitative research is harder to characterize in terms of the third factor: risk. Lubars' and Tan's conceptualization of risk is three-fold: personal accountability in case errors happen; the uncertainty, or the probability of errors; and the scope of impact, or cost or magnitude of those errors. These three components of qualitative research all focus on errors, but ``error'' is inherently difficult to define, if definable at all, in qualitative research to which subjective interpretations are central: What makes a researcher's interpretations \emph{wrong}? Is there such a thing as interpretation \emph{error}? Furthermore, the concept of risk itself may be more nuanced than currently characterized, given that Lubars and Tan did not find significant correlation between risk and delegability, and the complexity of defining ``error'' in qualitative research may also be present in other kinds of tasks. Designers should carefully consider this flexibility of qualitative research when designing AI assistance, especially supervised models that are entirely dependent on what is ``positive'' or ``negative.'' 

The final factor, trust, is also nuanced in the context of qualitative research. Trust, which Lubars and Tan also describe as containing three components, includes the perceived ability of AI to complete the task, the interpretability of AI's actions, and the perceived value alignment between AI and human. As our findings have shown, participants had low trust in AI's ability to reliably conduct qualitative analysis. While ``reliable'' is often defined as being consistently ``correct,'' our discussion above shows correctness may not be the focus in qualitative research since error itself is hardly definable. Instead, our findings suggest that the reliability lies in the consistent alignment of values---that the value of qualitative research derives from creativity and serendipity \cite{wiedemann_opening_2013}---between AI and human researchers, which our participants did not trust AI to have. P14 pointed out that she did not want any assistance because she could not ``unsee'' the AI suggestions, which indicates an assumption of unwanted, negative influence from AI that values concrete, highest-probability answers, as opposed to generative and nuanced ones sought by human-based qualitative analysis. However, influence is inevitable in collaborative qualitative research---having human collaborators also influences researchers' perspectives, and it is not reasonable to have to self-isolate to achieve the desired research integrity. Therefore, the assumed ``badness'' of AI influence might be because the suggestions come \emph{from} AI, rather than that the suggestions themselves are of low quality. 

The distrust of AI shared across our participants may possibly have to do with the differences between typical behavior of humans and AI. Humans usually make suggestions like ``This part of the data is interesting and maybe you should take a look at it'' (and more seasoned collaborators can also explain why it is interesting). However, AI suggestions, generated from classification results, often take the deterministic form of ``data X should be assigned code Y'' without explanations---even a human collaborator would be difficult to trust if they could only make such suggestions. While our participants did not explicitly mention AI's interpretability, the well-known opaqueness of the AI's inner working ``black box'' could only exacerbate the distrust.  P14's distrust echoes with Gach et al. 's finding that interpersonal trust---the willingness to accept risk based on the expectation of another person's behavior---often gets mistranslated when it is implemented as impersonal trust that a technical system will behave as expected \cite{gach_experiences_2020}. This mistranslation of trust resonates with Ackerman's notion of sociotechnical gap \cite{ackerman_intellectual_2000}---how can we better technically support the kind of interpersonal trust that we know we must support socially in qualitative research? 

Given the challenges related to risk and trust, a key consideration in the design of AI collaborative tools for qualitative research should be: How can we promote more interpersonal trust between human researchers and AI collaborators in ambiguous qualitative research? One way for AI to build trust is to provide humans with the ability to make choices and corrections \cite{knowles_models_2015}. If such ability is sufficient, one might think that an ideal collaboration would involve first having the AI automatically label the entire dataset, allowing humans to correct the AI as necessary afterwards. However, not only would such an approach meet the objections of participants (as exemplified by P14), Yan et al. \cite{yan_optimizing_2014} tried such an approach and concluded that creating a one-size-fits-all ML model for all codes in a multi-dimensional coding scheme was impossible. The low precision of their model suggests that humans may have to do \emph{more} work in correcting the AI than if they had coded on their own. While correction is also widely present in human collaboration, it is easier to discuss and resolve the differences with a human than with AI. Furthermore, it is also undesirable to have too much trust---if the data to be coded is ambiguous, our results show that it's possible that researchers might defer to AI as the \textit{de facto} tie breaker even though the situation deserves human deliberation to uncover the rich qualitative insights underlying that ambiguity. Given the two extremes, it may seem that offering answers that provide precisely the right balance of trust is the natural solution for AI, but ``just the right balance'' may never be practically achievable---the design of AI assistance for qualitative research may need to forgo the traditional answer-provider paradigm and leave the deliberation to humans.

\subsection{From Answers to Ambiguities and Disagreements}

How, then, can we preserve human researchers' agency in ambiguous situations in AI? A good starting point is to consider how it can play out in machine learning algorithms, a core component in many AI systems. Chen et al. \cite{chen_using_2018} point out that the goal of performance optimization in ML may be at odds with the goal of discovering patterns in qualitative research where the categories constantly change: while well-defined data categories are essential anchors for many machine learning algorithms, qualitative analysis does exactly the opposite by iteratively revising, challenging, and sometimes even overhauling existing categories to reveal deeper insights buried in the data; categories are not the end, but only a means to an end. Therefore, a machine learning model that identifies points of \emph{ambiguity} and \emph{disagreement} may have more utility to qualitative researchers. This point is further supported by McDonald et al., who discussed why seeking agreement might not be appropriate when codes are the  process instead of the product. This work also highlights the scenario where there is an expert researcher (or a researcher deemed as the ``expert'') in a team, in which case the other researchers defer to the expert researcher largely to maintain the social relationship. The expert-takes-all phenomenon was common among our participants, and we can speculate that it also broadly exists given how common advisor-student relationships and mentor-mentee relationships are in research training. McDonald et al. \cite{mcdonald_reliability_2019} also discussed how agreement can be harmful in research rooted in critical traditions (e.g., feminist HCI \cite{bardzell_feminist_2010}), and the type of ``forced'' agreement in this work may aggravate this harm. 

Taking the above considerations together, we can imagine an example assistive AI system that has the following features. First, to promote ambiguity and preserve human researchers' agency, the system could have the researcher choose the level of suggestions it provides. For example, the system could default to leaving the researcher on their own (i.e., do nothing) at first, and only later suggest interesting data snippets that the researcher has overlooked based on the researchers' existing coding/labeling patterns (e.g., through text classification using long short-term memory (LSTM) networks). The system could also suggest upcoming, unexamined data snippets for the researcher to consider, but only if the researcher specifically requests this. As demonstrated in our analysis, unprompted suggestions of new data can lead researchers into uncertain (and possibly unwanted) interpretive directions, a concern expressed by our participants. 

While it is possible to provide \emph{code} suggestions (likely based on topic modeling), none of our participants were enthusiastic about such an idea, ranging from hesitant to resistant. Our participants' reactions show an interesting contrast with the promise of machine learning-based coding shown in prior research \cite{baumer_comparing_2017, marathe_semi-automated_2018, chen_using_2018}. Considering these findings together, we urge AI designers to be careful in making suggestions about the exact codes or labels that a piece of data might take. An AI system capable of making code suggestions should be clear about what this feature is and is not (i.e., that instead of data to be coded, it makes suggestions \emph{of} codes). It should also be available only upon the researcher's request, and only after it learns enough of the researcher's coding patterns in a particular project.

Second, given that collaboration was common for many of our participants, the system could have features that further promote ambiguity and serendipity through collaboration. For example, it could suggest data snippets (examined or unexamined) that might be worthy of further discussion among collaborators. The system could also suggest a piece of data that a researcher has already coded to other collaborators who are likely to disagree with that researcher. Both examples aim to encourage discussions among researchers that may spark unexpected insights, and echo the point we made against careless coding suggestions---we envision the system to help set researchers up to do the research, instead of doing it for them.

Our vision above is only an example of what an assistive AI might look like, and the concrete specifications will depend on a number of factors such as the audience for whom it is designed (e.g., AI designed for HCI researchers vs. linguists are likely to be different), and the underlying infrastructure that it needs (e.g. collaborator-based suggestions will need collaboration functionality in the first place). The features that we envision also illustrate that an assistive AI will necessarily be an interactive one with mixed initiatives, some of which are already in line with principles of mixed-initiative designs \cite{horvitz_principles_1999}. While we acknowledge that such a system still does not address the expert-takes-all phenomenon, we are also hesitant to recommend that AI address the nuances of the social relationship between researchers. By revealing the the expert-takes-all phenomenon in the collaboration dynamics between junior and senior researchers, we are not calling for a technical solution to ``solve'' their relationship; we only hope that qualitative researchers can be aware of this phenomenon, and strive toward more productive collaborations with it in mind.

Uncertainty and ambiguity have long been viewed as a hindrance that costs efficiency and causes mistakes. However, uncertainty is a feature rather than a bug in qualitative research. As our findings indicate, while our participants wanted AI to help them have a better grasp of the uncertainty, they did not want it to be taken away. This very uncertainty allows qualitative researchers to dig deeply into their data, sometimes even to build personal connections with their data, and generate rich interpretations and insights. Our findings not only point to ways that AI can help qualitative researchers, but also reveal ways that qualitative researchers can help each other. For example, researchers who are not accustomed to collaboration may want to experiment with collaborating as a way to have various perspectives that promote serendipity. Senior researchers may also want to consider teaching junior researchers to embrace uncertainty as part of the qualitative research training, so that they can trust themselves in doing qualitative research on a high level while also maintaining a healthy dose of self-doubt.

\section{Limitations and Future Work}

Our research has several limitations. First, the majority of our participants were junior scholars, with arguably less experience in qualitative research than more senior ones. While we believe that a student majority is still able to provide valuable insights since they all actively conduct research, it may nevertheless have an impact on their reported experiences and tool use. 

Second, we would like to acknowledge that the qualitative research practices documented in this study (with the coding component in particular) are largely based on individuals, despite the ample descriptions of collaboration dynamics that we heard. We suspect the predominance of individual research practices is a product of the common student-advisor collaboration paradigm, and future work may find value in investigating more and other collaborative coding practices. Similarly, the kind of qualitative data that our participants dealt with were predominantly text, but the whole spectrum qualitative data is much broader, such as image, audio, and video, as some of our participants already indicated. We encourage future work to investigate these alternative forms of qualitative data and their analysis.

Finally, as we mentioned in the introduction, our work does not provide guidance for specific models or algorithms to be used. It is a first step toward successful human-AI collaboration in qualitative research by envisioning a world where AI assistance of qualitative research is possible, and laying out the promises and perils of that world. We hope that our findings will inform future in-depth studies of how specific AI systems can be adopted in qualitative research.

\section{Conclusion}
Qualitative inductive methods are widely used in CSCW and HCI research for their ability to generatively discover deep and contextualized insights, but these inherently manual and laborious processes are infeasible for analyzing large corpora, sacrificing the richness provided by these data. This study explores the potential of AI for assisting humans in conducting qualitative research by investigating current qualitative research practices and revealing where AI can come into play. We describe the methodological practices participants have, the insufficiencies of current tools to support these practices, and desired capabilities of their imaginary perfect tool. We also show that participants, when they collaborate, have complex and nuanced collaboration dynamics, and consensus reaching can be a negotiation of social relationships beyond the research project itself. Furthermore, while participants struggle with the messiness and uncertainty of qualitative analysis, they also want full agency of the process and insist that AI should not take it away from them. We argue that qualitative analysis is a unique case for human-AI collaboration because uncertainty is essential to qualitative analysis itself. The design of AI assistance should embrace this uncertainty and support serendipity by promoting human agency, instead of being the arbiter that aims to reduce uncertainty as AI has been traditionally conceptualized. 

\begin{acks}

We thank the reviewers for their significant time and care in improving this paper. We also thank our participants, many of whom sacrificed their conference time to participate in our study. Finally, we thank Brianna Dym, Steven Frost, Katie Gach, Shamika Goddard, Jacob Paul, Anthony Pinter, and Morgan Klaus Scheuerman for their help and feedback on the early drafts of this work.

This work was supported in part by the National Science Foundation Award \#1764089.

\end{acks}

\bibliographystyle{ACM-Reference-Format}
\bibliography{qbd}

%% file: participants-table.tex
\begin{table}
\begin{center}
\caption{Participant details.}
\label{tab:demographics}
\renewcommand{\arraystretch}{1.5}
\rowcolors{2}{lavenderblue}{white}
\begin{tabular}{ccc}
\rowcolor{white}
\textbf{Participant ID} & \textbf{Academic Position} & \textbf{Field of Study}      \\
\midrule
P01            & Y1 Ph.D. student    & Information Science \\
P02            & Y2 Ph.D. student    & Information Science \\
P03            & Y1 Ph.D. student    & Information Science \\
P04            & Y3 Ph.D. student    & Anthropology        \\
P05            & Y1 Ph.D. student    & Anthropology        \\
P06            & Y4 Ph.D. student    & HCI                 \\
P07            & Y6 Ph.D. student    & Information Science \\
P08            & Y3 Ph.D. student    & Information Science \\
P09            & Y3 Ph.D. student    & HCI                 \\
P10            & Y4 Ph.D. student    & Communication       \\
P11            & Postdoc           & HCI                 \\
P12            & Professor         & Anthropology        \\
P13            & Professor         & Communication       \\
P14            & Postdoc           & Information Science \\
P15            & Y6 Ph.D. student    & Linguistics         \\
P16            & Y3 Ph.D. student    & Linguistics         \\
P17            & Y4 Ph.D. student    & Journalism \\

\bottomrule
\end{tabular}
\end{center}
\end{table}

%% file: methods-table-1.tex
\begin{table}
\small
\begin{center}
\caption{Qualitative research methods used by participants. Here, we report the methods in participants' own words, so the same term may refer to different methods.}
\label{tab:methods}
\def\tabularxcolumn#1{m{#1}}
\rowcolors{2}{lavenderblue}{white}
\begin{tabularx}{\linewidth}{YYYYYY}
\rowcolor{white}
\textbf{Participant} & \textbf{Field of Study}                  & \textbf{Type of Data}                                              & \textbf{Data Collection Methods}                                                           & \textbf{Data Analysis Methods}                                                                                    & \textbf{Software}                \\
\midrule
P01         & Information Science             & Interview transcripts, API documentations                 & Semi-structured interview, content analysis                                       & Open Coding, ``seeing themes''                                                                               & Google Docs             \\
P02         & Information Science             & Interview transcripts                                     & Interview                                                                         & Qualitative coding, ``forming themes''                                                                       & MaxQDA                  \\
P03         & Information Science             & Interview transcripts, interview notes                    & Interview, note taking                                                            & Inductive coding, ``coming up with categories''                                                              & None                    \\
P04         & Anthropology                    & Field notes, voice recording, video recording             & Ethnography, participant observation, interview                                   & Coding, transcribing into International Phonetic Alphabet                                                                            & None                    \\
P05         & Anthropology                    & Interview transcripts, voice recording, field notes       & Participant observation, ethnographic interview, note taking                      & ``Highlight unusual things'', ``pulling up bigger themes''                                                       & Google Docs             \\
P06         & HCI                             & Online comments                                           & Scraping                                            & Thematic analysis                                                                                        & Yes, but did not remember \\
P07         & Information Science & Interview transcripts, field notes                        & Semi-structured interview, participant observation, expert interview, note taking & Transcribe, open coding, axial coding, memoing                                                           & None                    \\
P08         & Information Science             & Interview transcripts, field notes                        & Interview, participant observation                                                & Transcribe, code                                                                                         & ATLAS.ti                \\
P09         & HCI           & Interview transcripts, field notes                        & Interview, participant observation                                                & ``Classification process to organize data into themes''                                                      & Enable                  \\
\bottomrule
\end{tabularx}
\end{center}
\end{table}

%% file: methods-table-2.tex
\begin{table}
\small
\begin{center}
\renewcommand{\arraystretch}{1.2}
\def\tabularxcolumn#1{m{#1}}
\rowcolors{2}{lavenderblue}{white}
\begin{tabularx}{\textwidth}{YYYYYY}
\toprule
P10         & Communication                   & Interviews, field notes, eye tracking data                & Interview, experiment, eye tracking                                               & Transcribing, coding, affinity mapping                                                                   & ATLAS.ti                \\
P11         & HCI                             & Interview notes, interview excerpts                       & Semi-structured interview                                                         & Note taking, coding, ``seeing themes''                                                                       & Excel                   \\
P12         & Anthropology                    & Interview transcripts, field notes                        & Ethnography, participant observation, note taking, interview    & Content analysis, open coding, memoing, codebook \& closed coding, ``seeing themes''                                           & Dedoose                 \\
P13         & Communication                   & Interview transcripts                                     & Interview                                                                         & Transcribing, coding, ``coming up with coding schemes''                                                      & Word, Excel   \\
P14         & Information Science             & Interview transcripts                                     & Interview, ``kind of ethonographic stuff'', participant observation, participation  & Transcribing, micro-coding, ``inductively identifying themes'', ``grouping into categories using post-it notes'' & NVivo                   \\
P15         & Linguistics                     & Interview transcripts, clinical texts, online forum posts & Interview                                               & Public discourse analysis, coding, ``categorizing codes''                                                                               & MaxQDA                  \\
P16         & Linguistics                     & Video recordings                                          & Recording natural interaction                     & Transcribing into Jeffersonian system, ``making subcollections''                                             & Word                    \\
P17         & Journalism                      & Interview transcripts                                     & Interview, oral history                                                           & Discourse analysis                                                                                       & Excel, Google Docs \\
\bottomrule
\end{tabularx}
\end{center}
\end{table}